\documentstyle[epsf]{aipproc}
\def\H{Her X--1}   \def\B{BeppoSAX}
\begin{document}

\title{The pulsed light curves of \H\ as observed by \B}
\author{D. Dal~Fiume$^1$, M. Orlandini$^1$, G. Cusumano$^2$, \\
        S. Del~Sordo$^2$, M. Feroci$^3$, F. Frontera$^{1,4}$, \\
        T. Oosterbroek$^5$, E. Palazzi$^1$, A. N. Parmar$^5$, \\
        A. Santangelo$^2$, A. Segreto$^2$}
\address{$^1$ Istituto TESRE/CNR, via Gobetti 101, 40129 Bologna, Italy \\
 $^2$ Istituto IFCAI/CNR, via La~Malfa 153, 90146 Palermo, Italy \\
 $^3$ Istituto Astrofisica Spaziale/CNR, via Fermi 21, 00044 Frascati, Italy \\
 $^4$ Dipartimento di Fisica, Universit\`a di Ferrara, via Paradiso, 1, 44100 Ferrara, Italy \\
 $^5$ Astrophysics Division, Space Science department of ESA, ESTEC, P.O. Box 299, 2200 AG Noordwijk, The Netherlands
}
\maketitle

\begin{abstract}
We report on the timing analysis of the observation of the X--ray binary
pulsar \H\ performed during the \B\ Science Verification Phase. The
observation covered more that two full orbital cycles near the maximum of the
main--on in the 35 day cycle of \H. We present the pulse profiles from 0.1 to
100 keV. Major changes are present below 1 keV, where the appearance of a 
broad
peak is interpreted as re-processing from the inner part of the accretion disk,
and above 10 keV, where the pulse profile is less structured and the main peak
is appreciably harder. The hardness ratios show complex changes with pulse
phase at different energies.
\end{abstract}

\section*{INTRODUCTION}

\H\ is one of the most observed and best studied sources in the X--ray sky.
This eclipsing binary pulsar (orbital period 1.7 days, pulsation period 1.23
sec) was one of the first X--ray sources in its class to be discovered
\cite{843,847}. It shows a 35 day period on--off cycle in which a main--on and
a short--on are present. The flux from the source varies roughly a factor of
three between the main-on and the short--on \cite{583}. Low flux level emission
was also detected between the two on of the 35 day cycle.

The pulsed light curves observed during main--on show a broad,  structured
single peak from 2 to 100 keV \cite{1171}. There is evidence that the pulse
shape varies during the 35 day cycle both in the low energy band below 10 keV
and in the hard X--rays above 20 keV. In 1--30 keV a double peaked pulse shape
was observed with EXOSAT during the short--on state \cite{587}. Soong
et~al.\ \cite{1171}
measured with HEAO--1 a change in the 12--70 keV pulse shape during the
main on, even if they conclude that the changes they observe are more 
likely related to the source intensity than to the phase of the 35 day cycle.

\section*{OBSERVATION}

\B\ is a program of the Italian Space Agency (ASI) with participation of the
Netherlands Agency for Aerospace Programs (NIVR). It is composed by four
co-aligned Narrow Field Instruments (NFIs) \cite{1530}, operating in the energy
ranges 0.1--10 keV (LECS) \cite{1531}, 1--10 keV (MECS) \cite{1532}, 3--120 keV
(HPGSPC) \cite{1533} and 15--300 keV (PDS) \cite{1386}.  Perpendicular to the
NFI axis there are two Wide Field Cameras \cite{1534}, with a $40^\circ \times
40^\circ$ field of view.

During the \B\ Science Verification Phase (SVP) \H\ was observed from 1996 07
24 00:34:46 UT to 1996 07 27 11:54:46 UT. Data were telemetred in direct mode
for all the four NFIs, with information on arrival time, energy and position
for each photon. We report on the pulsed light curves observed during a
fraction of the out--of--eclipse phase of the binary orbit.

The data were recorded in single--event mode: each detected photon was
tagged with its arrival time in the detectors. The arrival times were
corrected to the solar system barycentre and folded at the pulsar period
of 1.2377396 s \cite{101}.
A correction for the Her X--1 orbital motion was also included.

\section*{RESULTS}

The folded light curves as observed by \B\ are shown in Figure 1. Two
clear transitions are seen: one at approximately 1 keV and the other at
approximately 10 keV.

The 1 keV transition goes from a broad sinusoidal pulse shape
(Figure 1, panel (a) ) to a more peaked pulse shape that remains 
almost unchanged up to 10 keV. The transition is accompanied by a phase 
shift of $\sim 250^\circ$.

The 10 keV transition goes from a structured single peak, with a prominent
shoulder on its trailing edge, to an almost perfectly triangular pulse
shape. The pedestal observed between phase 0.9 and phase 1.4 in Fig. 1
remains visible at least up to ~40 keV.
This second transition occurs exactly at the energy where the Her X--1
X--ray spectrum begins to deviate consistently from a simple power law
shape \cite{102}.

These two transition are even more evident in Fig. 2, in which we show
the ratios between light curves in different energy bands. The hardness
ratios also show some complex changes in the 2--10 keV energy interval.
These changes are percentually small, but are easily detected with high 
significance.

\bigskip

The transition at $\sim$1 keV is likely to be due to a change from a 
reflected to a directly observed pulse beam \cite{100,101}.
As discussed in \cite{101}, the phase shift of $\sim$250$^o$ observed 
between the soft and the hard pulse shape suggests that the reflection
zone is the inner part of a tilted accretion disk. 

The pulse width decreases with energy above 2 keV, mainly due to the
suppression at higher energies, above 10 keV, of the shoulder on the
trailing edge of the peak. This variation of the pulse shape
is likely due to a variation in the intrinsic beaming pattern with
energy, as predicted by emission models (e.g. \cite{104,103}).
However it is not straightforward to obtain the observed asymmetric shapes
directly from the models and from simple pencil or fan beams.

An attempt to introduce some degree of geometrical
complexity in the emission region was done by Panchenko and Postnov
\cite{105}, reproducing qualitatively the Her X--1 pulse shape.
However our detailed and simultaneous measurement in a
broad energy interval shows that any geometrically complex model must
explain also the energy dependence of the pulse shapes.
This can be done only by taking into account the
spatial anisotropy and energy dependence of the intrinsic beaming 
pattern from the emission zone.

\bigskip

\newlength{\figwidth}
\setlength{\figwidth}{\textwidth}
\addtolength{\figwidth}{0.1\textwidth}

\begin{figure}
\centerline{
\epsfxsize=\figwidth \epsfbox{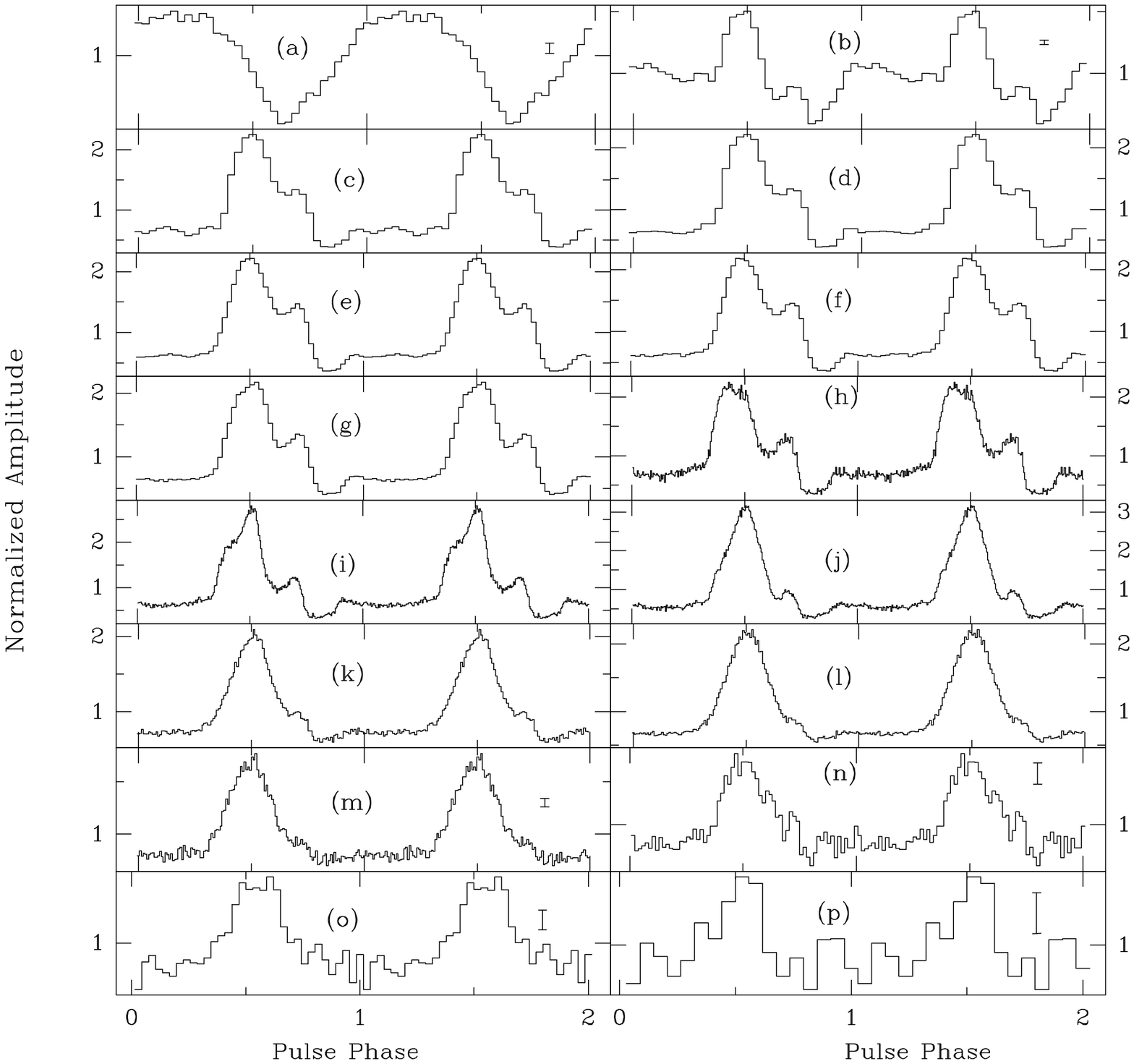}
}
\begin{caption}
{}
Folded light curves of Her X-1: (a) 0.1--0.4 keV, LECS; (b) 0.4--1.6
keV, LECS; (c) 1.6--2.4 keV, LECS; (d) 2.4--10 keV, LECS; (e) 2--4
keV, MECS; (f) 4--6 keV, MECS; (g) 6--10 keV, MECS; (h) 4--8 keV, HPGSPC;
(i) 8--15 keV, HPGSPC; (j) 15--30 keV, HPGSPC; (k) 13--20 keV, PDS; (l)
20--30 keV PDS; (m) 30--40 keV, PDS; (n) 40--50 keV, PDS; (o) 50--70
keV, PDS; (p) 70--100 keV, PDS. Where not indicated, the error bar is
smaller than the symbol size.
\end{caption}
\end{figure}

\begin{figure}
\centerline{
\epsfxsize=\figwidth \epsfbox{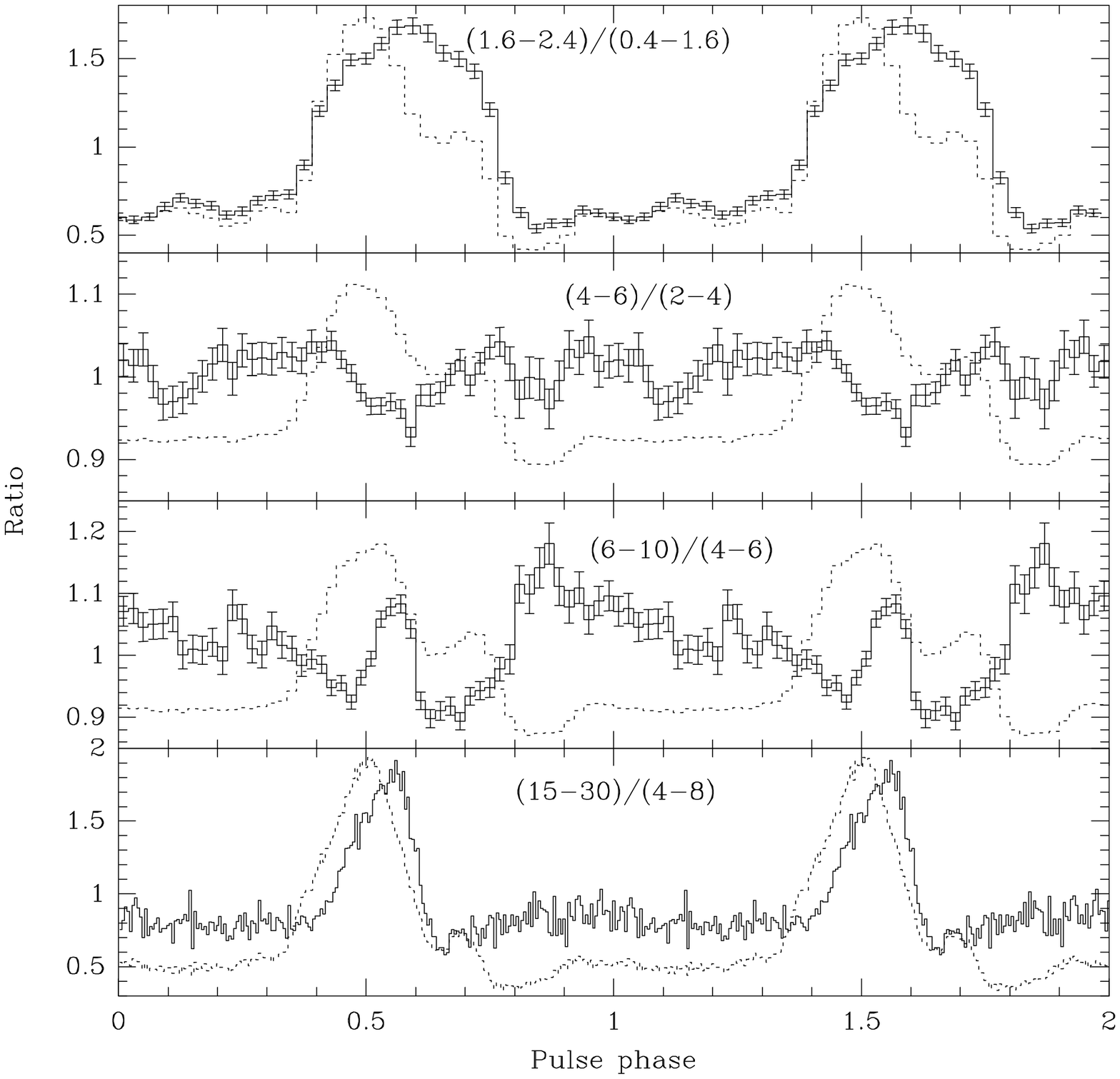}
}
\begin{caption}
{}
Hardness ratios between folded light curves in different energy bands.
In each panel
the dashed line shows the measured light curve in the higher energy band
for reference.
\end{caption}
\end{figure}

{\em Acknowledgements.\/}
The authors wish to thank the \B\ Scientific Data Center staff for their
support during the observation and data analysis. This research has been funded
in part by the Italian Space Agency.

\end{document}